\documentclass[%
 reprint,
showpacs,
footinbib,
 bibnotes,
 amsmath,amssymb,
 prl,
]{revtex4-1}

\usepackage{graphicx}
\usepackage{dcolumn}
\usepackage{bm}
\usepackage{color}
\usepackage{float}
\usepackage[colorlinks=true,linkcolor=blue,citecolor=blue]{hyperref}
\usepackage{multirow}
\usepackage{textcomp}
\usepackage{amsmath}

\begin{document}


\title{How flagellated bacteria wobble}
\author{Jinglei Hu$^{1}$}%
\email{hujinglei@nju.edu.cn}%
\author{Chen Gui$^{1\#}$}%
\author{Mingxin Mao$^{1\#}$}%
\author{Pu Feng$^{2\#}$}%
\author{Yurui Liu$^{1}$}%
\author{Xiangjun Gong$^{2}$}%
\email{msxjgong@scut.edu.cn}%
\author{Gerhard Gompper$^{3}$}
\email{g.gompper@fz-juelich.de}
\affiliation{$^1$Kuang Yaming Honors School, Nanjing University, Nanjing 210023, China}
\affiliation{$^2$Faculty of Materials Science and Engineering, South China University of Technology, Guangzhou 510640, China}
\affiliation{$^3$Theoretical Physics of Living Matter, Institute for Advanced Simulation, Research Center J\"{u}lich, 52425 J\"{u}lich, Germany}
%


%
\begin{abstract}
A flagellated bacterium navigates fluid environments by rotating its helical flagellar bundle. The wobbling of the bacterial body significantly influences its swimming behavior. To quantify the three underlying motions--precession, nutation, and spin, we extract the Euler angles from trajectories generated by mesoscale hydrodynamics simulations, which is experimentally unattainable. In contrast to the common assumption, the cell body does not undergo complete cycles of spin, a general result for multiflagellated bacteria. Our simulations produce apparent wobbling periods that closely match the results of {\it E. coli} obtained from experiments and reveal the presence of two kinds of precession modes, consistent with theoretical analysis. Small-amplitude yet periodic nutation is also observed in the simulations.

\end{abstract}

\maketitle

{\it Introduction}.---As one of the oldest living forms on Earth and a unique kind of active colloids, bacteria now are also studied as models for non-equilibrium systems and motile active matter. Peritrichous bacteria such as {\it Escherichia coli} exploit a bundle of multiple helical flagella for locomotion~\cite{Book-2004-Berg}. Numerous studies have been performed to unravel the swimming behavior of single {\it E. coli} or other flagellated bacteria in bulk~\cite{JBacteriol-2007-Darnton, BJ-2010-Larson, PNAS-2014-Martinez, PNAS-2014-Liu, SoftMatter-2015-Hu, SciRep-2015-Patteson, NatPhys-2019-Zottl, Nature-2022-Kamdar} or near surfaces~\cite{Nature-2005-DiLuzio, BJ-2006-Lauga, PRL-2008-Lauga, PRL-2011-Leonardo, PRL-2014-Molaei, SciRep-2015-Hu, PRX-2017-Bianchi, PNAS-2022-Cao}, as well as the collective behavior of bacterial swarms~\cite{PRL-2012-Chen, Nature-2017-Chen, Nature-2021-Liu}. The wobbling motion of the cell body around its swimming direction was initially visualized for {\it E. coli}~\cite{JBacteriol-2007-Darnton}, and has been experimentally shown to profoundly affect the bacterial swimming behavior. For instance, wobbling induces helical swimming paths for {\it Bacillus subtilis}~\cite{JFluidMech-2012-Hyon}, {\it Caulobacter crescentus}~\cite{PNAS-2014-Liu}, and {\it Helicobacter pylori}~\cite{SciAdv-2016-Maira}. The suppression of wobbling contributes to the enhanced swimming speed of {\it E. coli} in complex fluids~\cite{SciRep-2015-Patteson, Nature-2022-Kamdar}. However, the kinematics and dynamics of bacterial wobbling remain elusive.

The wobbling motion detected in experiments~\cite{JBacteriol-2007-Darnton, JFluidMech-2012-Hyon, PNAS-2014-Liu, PRE-2015-Bianchi, SciAdv-2016-Maira} reflects the precession of the bacterial body that has an off-axis orientation with respect to the counterrotating flagellar bundle. As a rotating rigid body, the bacterial body might nutate while precessing. This naturally leads to fundamental questions, like: How does the bacterium precess, with one direction or with two alternating directions? Does the bacterium exhibit periodic nutation? If so, what is the rate of nutation compared to precession?

Our comprehension of the bacterial wobbling kinematics and dynamics is limited by experimental challenges in tracking the orientations of the body and the flagellar bundle. Phase-contrast microscopy~\cite{JBacteriol-2007-Darnton, SciAdv-2016-Maira, Langmuir-2023-Zheng} offers direct 2D images of the whole bacterium, while digital holography~\cite{PNAS-2012-Su, PRL-2014-Molaei, PNAS-2014-Liu, Langmuir-2017-Qi, PRX-2017-Bianchi} provides 3D trajectories of bacterial swimming across a large volume by reconstructing the coordinates of the cell body's centroid. Though apparent wobbling periods can be derived from these images and trajectories, discerning wobbling kinematics—such as precession mode, nutation, and spin—requires knowledge of orientations of both the body and the flagellar bundle, as will become evident through our detailed analysis below, which is not accessible experimentally. From a theoretical standpoint, the friction between the cell body and the multiple flagella as well as near-field hydrodynamics~\cite{SoftMatter-2015-Hu} pose substantial challenges for an analytical description of the wobbling of peritrichous bacteria.

In this Letter, we overcome the experimental and analytical limitations by performing extensive hydrodynamics simulations of a quantitative mechanical model~\cite{SciRep-2015-Hu, SoftMatter-2015-Hu, SoftMatter-2020-Mousavi} of {\it E. coli}, and decipher the intricate wobbling kinematics through detailed analysis of the resultant trajectories in combination with theoretical analysis. In our simulations, short polymer chains which experience purely repulsive interactions with the bacteria are added to the solvent to tune the average angle between the body and the flagellar bundle, i.e., to adjust the amplitude of the wobbling. These simulations produce apparent wobbling periods that closely match the results of {\it E. coli} obtained from experiments by Darnton et al.~\cite{JBacteriol-2007-Darnton} and by ourselves. From the orientations of the body and the flagellar bundle tracked over the course of simulations, the Euler angles for the rotation of the cell body were extracted to quantify the three types of motions underlying the apparent wobbling--precession, nutation and spin. Theoretical analysis contributes to a further understanding of the wobbling kinematics.

\begin{figure*}[tbh]
\includegraphics[width=2\columnwidth]{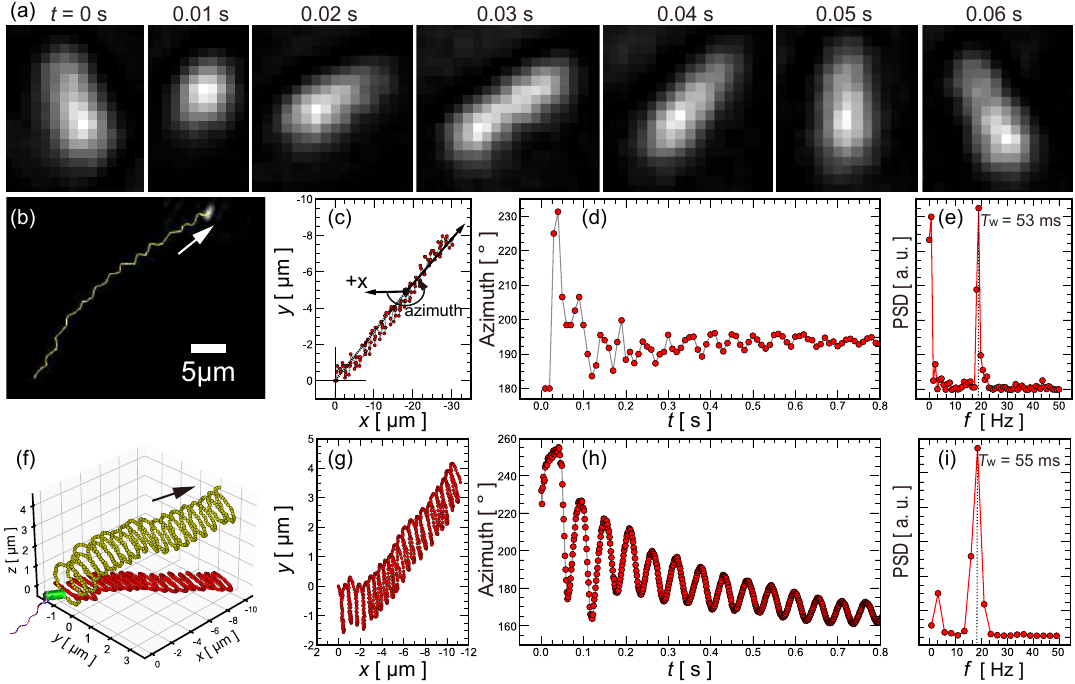}
\caption{Analysis of {\it E. coli} wobbling using experimental (a-e) and simulated (f-i) swimming trajectories. (a) Sequential 2D optical images of the bacterial body, where each pixel spans $0.1625 \times 0.1625$ $\mu$m$^2$. (b) 2D trajectory constructed using the coordinates of the brightest pixels in consecutive frames as shown in (a). The arrow indicates the swimming direction. (c) 2D trajectory displayed in the $xy$ plane, where each point defines an azimuth angle. (d) Time series of the azimuth for the trajectory in (c). Only the initial part is shown here for better visualization of the oscillation. (e) Power spectrum of the azimuth in (d). (f) 3D helical trajectory of a tracer particle located at one pole of the body from simulations. The red trajectory is a projection of the helical trajectory onto the $xy$ plane. (g)-(i) are from the trajectory in (f), and correspond to (c)-(e), respectively.}
\label{fig:1}
\end{figure*}

{\it Results and discussion}.---Figure~\ref{fig:1}a shows a time sequence of 2D images of the {\it E. coli} body reconstructed in the focal plane using digital holographic microscopy (see the Supplementary Materials for details). The body restores its original orientation ($t = 0$) in around 60 ms, suggesting a wobbling motion with about the same period. A 2D swimming trajectory (Fig.~\ref{fig:1}b) can then be constructed from the coordinates of the brightest pixels in such consecutive images. To measure the apparent wobbling period more precisely, an azimuth angle is defined for each point of the trajectory in the $xy$ plane (Fig.~\ref{fig:1}c). The oscillation of the azimuth (Fig.~\ref{fig:1}d) reflects the body's wobbling evident in the swimming trajectory. The power spectrum of the azimuth (Fig.~\ref{fig:1}e) exhibits a primary peak with a period of $T_\text{w} = 53$ ms, close to our previous estimate of 60 ms and Darnton {\it et al.}'s experimental result of 80 ms~\cite{JBacteriol-2007-Darnton}. Simulation results for the model {\it E. coli}, with four asymmetrically distributed flagellar anchors and swimming in a polymer solution of concentration $c = 0.15$ (see SM for details), are displayed in Figs.~\ref{fig:1}f-\ref{fig:1}i. The 3D yellow trajectory of a tracer bead at one pole of the body (Fig.~\ref{fig:1}f) illustrates a helical swimming path, as observed for many flagellated bacteria~\cite{JFluidMech-2012-Hyon, PNAS-2014-Liu, SciAdv-2016-Maira}. The projection of the 3D trajectory onto the $xy$ plane, using the same procedure as for the experimental data in Fig.~\ref{fig:1}c-\ref{fig:1}e, yields a period of wobbling $T_\text{w} = 55$ ms in Fig.~\ref{fig:1}i consistent with the experimental result of $53$ ms. Although the experimental and simulated {\it E. coli} swim in fluids of different viscosities and are likely to possess different angles between the body and the flagellar bundle, the agreement supports the validity of our simulations. The apparent wobbling period of $T_\text{w}$ corresponds to the period of the body's precession, as will be shown in Fig.~\ref{fig:3}b.

\begin{figure*}[tbh]
\centering
\includegraphics[width=2\columnwidth]{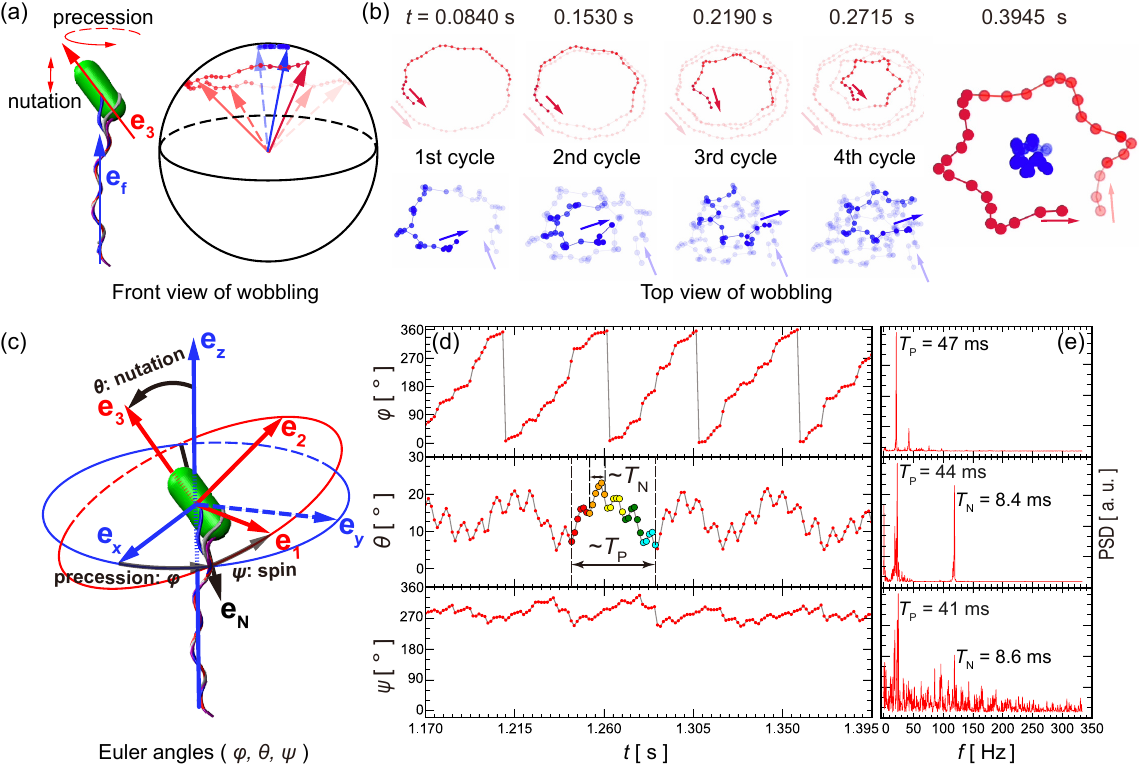}
\caption{Kinematic analysis of the wobbling of simulated {\it E. coli} with asymmetric flagellar anchors at $c = 0.20$. (a) Front view of the wobbling. The precession and nutation of the spherocylindrical body or the helical flagellar bundle are tracked {\it via} the trajectory of one end of its axis (${\bf e}_3$ or ${\bf e}_\text{f}$) on a unit sphere's surface, with the other end fixed at the sphere's center. (b) The initial four cycles of trajectories traced by ${\bf e}_3$ (in red) and ${\bf e}_\text{f}$ (in blue) are depicted in a top view, although not drawn to scale for clarity. Both the body and the flagellar bundle precess counterclockwise. Trajectories in the rightmost panel are accurately scaled. (c) Two coordinate systems defined by the orthogonal bases $({\bf e}_x, {\bf e}_y, {\bf e}_z)$ and $({\bf e}_1, {\bf e}_2, {\bf e}_3)$ are used to derive the Euler angles $(\varphi,\,\theta,\,\psi)$ that quantify the precession, nutation, and spin of the body. See the main text for details. (d) Time series of the Euler angles extracted from the simulation data. (e) Power spectra of the angles in (d).}
\label{fig:2}
\end{figure*}

We proceed with the detailed analysis of the wobbling based on simulations. As shown in Fig.~\ref{fig:2}a, since the precession and nutation of the spherocylindrical body or the helical flagellar bundle cause the reorientation of its axis (${\bf e}_3$ or ${\bf e}_\text{f}$), these two kinds of motions can be visualized by tracking the trajectory of one end of the axis on the surface of a unit sphere, while fixing the other end at the sphere's center. Figure~\ref{fig:2}b provides a top view of the initial four cycles of the trajectories swept by ${\bf e}_3$ (in red) and ${\bf e}_\text{f}$ (in blue). The trajectories are not drawn to scale for clarity. Both the body and flagellar bundle precess counterclockwise~\footnote{The overall rotation of the body is primarily due to precession, while the overall rotation of the flagellar bundle is dominated by spin. These two overall rotations are in opposite directions.}, and complete nearly an equal number of cycles within the same time frame, implying synchronization in their precession. This finding explains the experimental observation that the orientation of {\it Helicobacter pylori}'s flagellar bundle changes with respect to the body~\cite{SciAdv-2016-Maira}. The nutation of the body and the flagellar bundle becomes apparent starting from the third cycle and is also synchronized, as evidenced by the concurrent emergence of distinct `corners' in the trajectories. One cycle of accurately scaled trajectories in the steady state is provided in the rightmost panel of Fig.~\ref{fig:2}b, each comprising 32 points with a total duration of $46.5$ ms. There are six `corners' in the single cycle of each trajectory, indicating that the precession period $T_\text{P}$ is about 6 times the nutation period $T_\text{N}$. We then have the estimates $T_\text{P} \approx 46.5$ ms and $T_\text{N} \approx 7.75$ ms. The observation that ${\bf e}_3$ traces a much larger cycle than ${\bf e}_\text{f}$ motivates us to regard ${\bf e}_\text{f}$ as the precession axis ${\bf e}_z$ of the body for kinematic analysis.

\begin{figure*}[tbh]
\centering
\includegraphics[width=2\columnwidth]{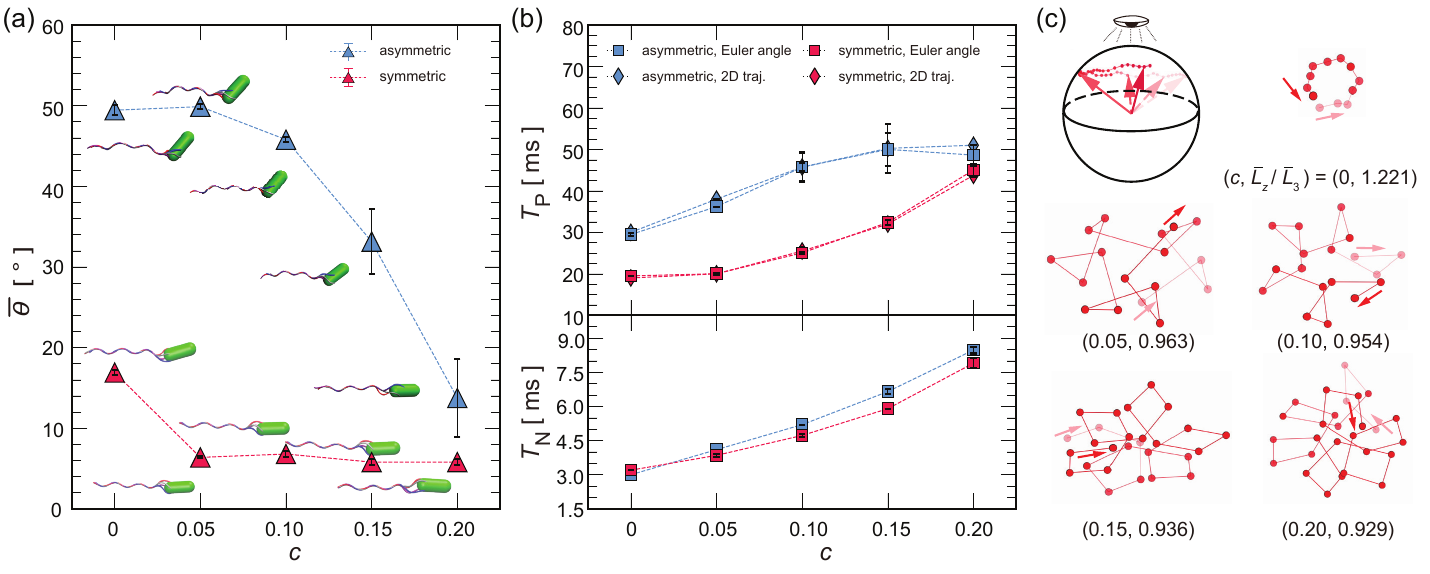}
\caption{(a) Average angle $\bar{\theta}$ between the body's long axis and the flagellar bundle's axis {\it versus} concentration $c$ of the polymer solution for {\it E. coli} with asymmetric or symmetric flagellar anchors. Inserted are simulation snapshots of the bacteria. (b) Period of precession $T_\text{P}$ and nutation $T_\text{N}$ of the body {\it versus} $c$. $T_\text{P}$ is derived from either 2D trajectories or Euler angles as indicated by the legend. $T_\text{N}$ is extracted from Euler angle $\theta$. (c) Trajectories similar to Fig.~\ref{fig:2}b but for {\it E. coli} with symmetric flagellar anchors. The last four trajectories are zoomed in 4$\times$ to better visualize the precession with alternating directions. At each value of $c$, the ratio $\bar{L}_z / \bar{L}_3$ is determined from simulation averages.}
\label{fig:3}
\end{figure*}

To quantify the wobbling kinematics, we extract the Euler angles of the bacterial body from its long axis ${\bf e}_3$ and the flagellar bundle's axis ${\bf e}_z$ monitored throughout the simulations. It remains  a challenge to acquire the two axes through laboratory experiments. Two coordinate systems are used to derive the Euler angles $\varphi$, $\theta$, and $\psi$ for the description of the precession, nutation, and spin of the body, as shown in Fig.~\ref{fig:2}c. The flagellar bundle determines the reference coordinate system $({\bf e}_x, {\bf e}_y, {\bf e}_z)$, whereas the other orthogonal basis $({\bf e}_1, {\bf e}_2, {\bf e}_3)$ specifies the body-fixed coordinate system. ${\bf e}_\text{N}$ defines intersection line of the ${\bf e}_x$-${\bf e}_y$ and ${\bf e}_1$-${\bf e}_2$ planes. $\varphi$ is then defined by ${\bf e}_x$ and ${\bf e}_\text{N}$, $\theta$ by ${\bf e}_z$ and ${\bf e}_3$, and $\psi$ by ${\bf e}_\text{N}$ and ${\bf e}_1$. The time series of the three angles in Fig.~\ref{fig:2}d exhibits periodic variations. Surprisingly, the spin angle $\psi$ fluctuates within a narrow range instead of varying monotonically between $0^\circ$ and $360^\circ$, indicating that the body undergoes slight back-and-forth rotation around ${\bf e}_3$ without a distinct direction. This finding is further supported by the unnoticeable change in the tracer's position relative to the body's axis during precession (Fig.~S2a), and contradicts the common assumption that the bacterial body in general rotates around its own axis in the opposite direction to the rotation of the flagellar bundle. For multiflagellated bacteria, the body and the flagellar bundle precess simultaneously through constant adjustments of each flagellum's anterior part (Movie S1). However, complete cycles of body spin, which would cause continuous twisting and stretching of each flagellum, are physically impossible~\footnote{There are no such physical constraints in uniflagellated bacteria. The spin of body at a rate of about 30 Hz has been reported for {\it Caulobacter crescentus}, when the single flagellum pushes the cell forward~\cite{PNAS-2014-Liu}.}. Conversely, the flagellar bundle spins in the opposite direction to the precession at a rate about 5 times faster than the precession (Fig.~S2b-d and SM text).

The power spectra in Fig.~\ref{fig:2}e for the Euler angles unveil the periods of the underlying motions. The upper panel for the precession angle $\varphi$ reveals a period of $T_\text{P} = 47$ ms, consistent with the estimate of $T_\text{P} \approx 46.5$ ms from visual inspection of the trajectories in the rightmost panel of Fig.~\ref{fig:2}b and the apparent wobbling period $T_\text{w} = 48$ ms derived from the 2D swimming trajectory (Fig.~S3). The middle panel for the nutation angle $\theta$ exhibits a secondary peak with a period of $T_\text{N} = 8.4$ ms, in accord with the previous estimate of nutation period $T_\text{N} \approx 7.75$ ms. This signal of nutation period clearly shows that the nutation is driven by an active torque instead of thermal noise. A closer look at the time series of $\theta$ in Fig.~\ref{fig:2}d reveals that one precession cycle roughly consists of 29 points, with a total duration of 42 ms. The high-frequency periodic fluctuations of about $5^\circ$ around the local average of $\theta$, as depicted by the five groups of points in different colors, mirror the body's small-amplitude nutation. Such nutation is hardly detectable from power spectral analysis of the 2D swimming trajectory with high spatial and temporal resolutions (SM text and Fig.~S3). Experimental measurement of the nutation therefore must be exceedingly difficult. The quantitative agreement between the power spectral analysis and the visual inspection of both the time series and the trajectories in Fig.~\ref{fig:2} confirms the reliability of our approach for quantifying the precession and nutation using Euler angles. The emergence of the precession signal ($T_\text{P} = 44$ ms) in the power spectrum of the nutation angle $\theta$ and the emergence of precession ($T_\text{P} = 41$ ms)  and nutations ($T_\text{N} = 8.6$ ms) signals in the power spectrum of the spin angle $\psi$ are explained in detail in the SM text.

Figure~\ref{fig:3}a shows that the average nutation angle $\bar\theta$ can be tuned by varying the concentration $c$ of the polymer solution. Here, $\theta$ is defined by the body's long axis and the flagellar bundle's axis, as depicted in Fig.~\ref{fig:2}c. As $c$ increases, the polymer-bacterium hardcore repulsion tends to restrict the wobbling of the body, leading to a decrease in $\bar\theta$, similar to experimental observations in ref.~\cite{SciRep-2015-Patteson}. Figure~\ref{fig:3}b shows that the period of precession $T_\text{P}$ and nutation $T_\text{N}$ of the body increase with $c$, primarily due to the slowdown of rotations caused by increased viscosity (Fig.~S1). The values of $T_\text{P}$ extracted from the 2D swimming trajectories (diamonds) agree with those from Euler angles (squares). The ratio $T_\text{P} / T_\text{N}$ indicates that the nutation is about 5--9 times faster than the precession. Since the bacterium swims in a low-Reynolds-number regime, the nutation of the bacterial body is driven by an active torque of the same period. This active torque arises from the flagellum-body friction~\cite{SoftMatter-2015-Hu} and from the torques that counteract each flagellum's rotary motor. The nearly identical values of $T_\text{N}$ at the same polymer concentration $c$ for {\it E. coli} with symmetric or asymmetric flagellar anchors suggest that the period of this active torque seems rather insensitive to the flagellar anchoring.

Figure~\ref{fig:3}c illustrates the two types of precession modes for {\it E. coli} with symmetric flagellar anchors. In pure solvent ($c = 0$), the cell body exhibits normal precession with counterclockwise direction when viewed from the top. In polymer solutions ($c = $ 0.05 -- 0.20), the body precesses with alternating directions, i.e., short-term clockwise and long-term counterclockwise. The two modes can be understood by considering the precession angular velocity $\dot\varphi$. In the steady state, the two components of the angular momentum ${\bf L}$ of the body, $L_3 = {\bf L} \cdot {\bf e}_3 = I_3 (\dot{\psi} + \dot{\varphi} \cos \theta)$ and $L_z = {\bf L} \cdot {\bf e}_z = (I_1 \sin^2 \theta + I_3 \cos^2 \theta) \dot{\varphi} + I_3 \dot{\psi} \cos \theta$, can be considered as constant quantities~\footnote{Our mesoscale hydrodynamics simulations which automatically include thermal fluctuations reveal that the time averages of the two instantaneous quantities, $\bar{L}_z$ and $\bar{L}_3$, are constants.}. $I_1$ and $I_3$ are the moments of inertia of the spherocylindrical body along ${\bf e}_1$ and ${\bf e}_3$, respectively. The precession angular velocity is then
\begin{equation}
\dot{\varphi} = (L_z - L_3 \cos\theta)/I_1 \sin^2\theta.
\label{eq:dotphi}
\end{equation}
Equation~(\ref{eq:dotphi}) implies that normal precession ($\dot\varphi > 0$) occurs when $L_z > L_3$. When $L_z < L_3$, there exists a value of $\theta$ at which $\dot\varphi = 0$. In this case, precession with alternating directions occurs. The two precession modes of {\it E. coli} discovered in our simulations perfectly match the predictions from Eq.~(\ref{eq:dotphi}) (Fig.~S4). It is worthy to note that the friction is adsorbed in the two angular momenta $L_z$ and $L_3$. The stronger the friction, the smaller these two quantities.

The spin angular velocity can also be derived as $\dot{\psi} = L_3 / I_3 -  \dot\varphi \cos\theta  = L_3 / I_3 -  L_3 / I_1 + (L_3 - L_z \cos\theta) /I_1 \sin^2\theta$. The Taylor expansion of the two $\theta$-functions around the average nutation angle $\bar\theta$ elucidates the emergence of the signals of precession, nutation, and their coupling in the power spectrum of $\psi$ (lower panel of Fig.~\ref{fig:2}e). For details see SM text.

{\it Conclusions}.---We have unraveled the wobbling kinematics of the bacterium {\it E. coli} through an investigation combining experiments, simulations, and theoretical analysis. The apparent wobbling period measured from both experimental and simulated swimming trajectories corresponds essentially to the precession period. In addition, the two precession modes discovered in our simulations are consistent with the theoretical analysis. Surprisingly, the nutation is barely detectable even with highly detailed simulated swimming trajectories surpassing experimental resolution, and only becomes evident by tracing the evolution of orientations of body and flagellar bundle. The nutation is found to be 5--9 times faster than the precession. Our study provides insights into the kinematics of bacterial motion at the single-cell level, and contributes to our understanding of nonequilibrium physics in active systems.

This work was supported by National Natural Science Foundation of China Grants No. 21973040 (J. H.) and No. 21973032 (X. G.), and by Fund of Guangdong Provincial Key Laboratory of Luminescence from Molecular Aggregates at South China University of Technology Grant No. 2023B1212060003 (X. G.).


%

\end{document}